\documentclass[letterpaper]{article}
\usepackage{graphicx}
\begin{document}
\title{Generalization of Equatorial Impact-Parameter Formulas for Rotating Bodies}

\author{James Graber\footnote{jgra@loc.gov} \\407 Seward Square SE  \\ Washington, Dc 20003   }

\maketitle

\begin{abstract}
This paper computes co-rotating and contra-rotating impact-parameter formulas in the plane of symmetry 
for any plane symmetric and 
axisymmetric rotating body in all metric theories of gravity, including general relativity.  
Impact-parameter formulas are useful to compute the appearance of accreting black holes, neutron stars, 
and other emitting or reflecting matter 
near a gravitationally compact rotating body.
These rotating-body impact-parameter formulas generalize similar impact-parameter formulas for the Kerr case 
derived by Bardeen and coworkers in 1972, 
and another general-metric formula for the spherical case published by Bodenner and Will in 2003.

\end{abstract}

\section{Introduction}
The impact parameter, also called the radius at infinity, is a coordinate-system-independent value for the position at which a 
light ray passing near a compact body is seen.  This formula for calculating the impact parameter is useful for computing the 
gravitational deflection of light, other gravitational lensing effects, and the appearance of black holes and neutron stars. 
It is particularly useful for predicting the appearance of an emitter or a reflector orbiting a black hole, and hence the 
apparent diameter of the black hole. The impact-parameter formula depends on the position of the emitter, 
the mass of the compact body, and the metric, which in effect describes both the compact body and the theory of gravity being used.  
Usually the compact body is a black-hole candidate, and the theory of gravity is general relativity.  Recently, however, 
a surprisingly small result for the apparent diameter of Sgr A* [1], the black-hole candidate at the center of the Milky Way galaxy, 
has created renewed interest in calculating apparent diameters in alternate theories of gravity.  This paper presents the 
impact-parameter formulas for co-rotating and contra-rotating observations in the equatorial plane of a symmetric rotating body 
in any metric theory of gravity.  It generalizes previous equatorial rotating-body results from the Kerr case in general relativity, 
originally computed by Bardeen \textit{et al.} [2,3], to the corresponding case in any metric theory of gravity.   It also 
generalizes spherical-case results for general spherical metrics recently published by Bodenner and Will [4].

\section{History}
Einstein first considered gravitational deflection of light in 1908 [5], long before he completed general relativity in 1915.  
Einstein computed the impact-parameter formula for the Schwarzschild solution to the first order in the inverse radius in 1916 [6], 
and the formula was used to predict the results of Eddington's 1919 [7,8] eclipse test of the gravitational deflection of light.  
Bardeen and coworkers computed the impact-parameter formula for the Kerr solution, including the equatorial case in 1972 [2,3].  
Bodenner and Will published the impact-parameter formula for a general spherical metric in 2003 [4], generalizing the Schwarzschild case.  
This paper, which presents the impact-parameter formula in the equatorial plane of a general rotationally symmetric metric of the Kerr type, 
generalizes both the equatorial-plane rotating-case results of Bardeen \textit{et al.}, and the spherical-case results of Bodenner and Will.

\section{Notation}
The impact-parameter formula relates the radius at infinity of a null geodesic in an asymptotically flat quasi-spherical metric space 
to the radius of closest approach.  In effect, it allows one to compute a coordinate-system-free, physically observable value of the 
apparent radius of an object near a black hole or a neutron star.  It depends on the mass and rotation of the central object, 
as well as the metric (coordinate system) describing the central body, and the metric-dependent radius of closest approach.

Following traditional notation, we will use $b$ for the metric-free impact parameter,  $r_0$ for the metric-dependent radius of closest approach, 
$M$ for the mass of the central object and  $J$ (or $aM$  where  $a =J/M$)for the angular momentum of the central body.  
The metric will be expressed in terms of the usual $r,\theta  ,\phi$   spherical coordinates.

\section{Comparison of Results}
The classic deflection-of-light result is usually computed in the standard or isotropic Schwarzschild metric and stated as

\begin{eqnarray}
\Delta \,\phi  = \frac{-4\,M}{b}
\end{eqnarray}

where $b$ is the impact parameter, $M$ is the mass of the deflecting body and 
${\Delta \phi }$ is the light deflection.  To zeroth order, $b=r_0$.

The standard Boyer-Lindquist cordinates for the Kerr metric simplify in the equatorial plane to

\begin{eqnarray}
{{ds}}^2 = \frac{-2\,{dt}\,
      {d\phi }\,{J}\,M}{r} + 
   \frac{{{dt}}^2\,
      \left( -2\,M\,r + r^2 \right) }{r^2} + 
   \frac{{{dr}}^2\,r^2}
    {{{J}}^2 - 2\,M\,r + r^2} + 
   {{d\phi }}^2\,
    \left( r^2 + \frac{{{J}}^2\,
         \left( 2\,M + r \right) }{r} \right)
\end{eqnarray}

which leads to the following equatorial-plane impact-parameter formula, originally computed by Bardeen \textit{et al.} and 
recently used by Broderick, Loeb and Narayan[9] to analyze the Sgr A* observations. 

\begin{eqnarray}
b = \frac{2\,{J}\,M \pm 
     r_0\,{\sqrt{{{J}}^2 + 
          r_0\,\left( -2\,M + r_0 \right) }}}{2\,M - r_0}
\end{eqnarray}

The metric used by Bodenner and Will is 

\begin{eqnarray}
{{ds}}^2 = - {{dt}}^2\,A(r)
        + {{dr}}^2\,B(r) + 
   \left( {{d\theta }}^2 + 
   {{d\phi }}^2\,{{sin}}^2\,\theta \right)\,r^2\,{C}(r)
\end{eqnarray}

Their impact parameter result is

\begin{eqnarray}
b = {r_0}\,{\sqrt{\frac{{C}(
         {r_0})}{A({r_0})}}}
\end{eqnarray}

The simplified metric we use in the plane of symmetry is 

\begin{eqnarray}
{{ds}}^2 = - dt^2\,A(r)   + 
   {{dr}}^2\,B(r) +   
   {dt}\,{d\phi }\,
    {F}(r) + 
   {{d\phi }}^2\,{C}(r)\,
    {\sin (\theta )}^2
\end{eqnarray}

Our result is 

\begin{eqnarray}
b = \frac{F(r_0) + {\sqrt{A(r_0)\,C(r_0) + F(r_0)^2}}}
   {A(r_0)}
\end{eqnarray}

for the contra-rotating case,  

and

\begin{eqnarray}
b = \frac{F(r_0) - {\sqrt{A(r_0)\,C(r_0) + F(r_0)^2}}}
   {A(r_0)}
\end{eqnarray}

for the co-rotating case.  



\section{Outline of Derivation}
We begin with a totally general line element for a rotating body with a central plane of symmetry

\begin{eqnarray}
{{ds}}^2 = - {{dt}}^2\,
      A(r,\theta )   + 
   {{dr}}^2\,B(r,\theta ) + 
   {{d\theta }}^2\,
    {D}(r,\theta ) + 
   {dt}\,{d\phi }\,F(r,\theta ) + 
   {{d\phi }}^2\,{C}(r,\theta )\,
    {\sin (\theta )}^2
\end{eqnarray}

Step 1:
We simplify the expression for the metric in the plane of symmetry of the rotating body, and following Bardeen, write the standard geodesic equation 

\begin{eqnarray}
-{\mu}^2 ds^2=  - dt^2\,A(r)   + 
   {{dr}}^2\,B(r) +   
   {dt}\,{d\phi }\,
    {F}(r) + 
   {{d\phi }}^2\,{C}(r)\,
    {\sin (\theta )}^2
\end{eqnarray}

Here,  $(dr/ds, d \theta /ds, d\phi/ds, dt/ds)$ is the tangent vector describing the geodesic, $\mu$ is 
the rest mass of the test particle following the geodesic, and $s$ is the usual affine parameter.

Step 2:  Following Bardeen we substitute two constants of motion  
($E$, the energy relative to infinity, and $\Phi$, the angular momentum about the symmetry axis) and solve for $(dr/ds)^2$

\begin{eqnarray}
\left( \frac{dr}{ds}\right)^2 = 
  \frac{-\left( {\Phi }^2\,A(r) \right)  - 
     \left( -{{E}}^2 + {\mu }^2\,A(r) \right)
        \,{C}(r) + 
     2\,{E}\,\Phi \,F(r) - {\mu }^2\,{F(r)}^2}
     {A(r)\,B(r)\,{C}(r) + B(r)\,{F(r)}^2}
\end{eqnarray}

Step 3: We next specialize to the photon case by setting $\mu = 0$  and re-expressing the above equation 
in terms of the impact parameter, $b$, by substituting $\Phi = b E$

\begin{eqnarray}
 0 = -\left( \frac{r^2\,\left( b^2\,{{E}}^2\,
          A(r) - {{E}}^2\,{C}(r) - 2\,b\,{{E}}^2\,F(r) \right) }{A(r)\,
        B(r)\,{C}(r) + B(r)\,{F(r)}^2} \right)
\end{eqnarray}

Step 4: Finally, we solve for the impact parameter in the co-rotating case

\begin{eqnarray}
b = \frac{F(r_0) - {\sqrt{A(r_0)\,C(r_0) + F(r_0)^2}}}
   {A(r_0)}
\end{eqnarray}

and in the contra-rotating case

\begin{eqnarray}
b = \frac{F(r_0) + {\sqrt{A(r_0)\,C(r_0) + F(r_0)^2}}}
   {A(r_0)}
\end{eqnarray}

\section{Conclusion}
The results are two formulas for the impact parameter in the equatorial plane that are conveniently compact.  
The generalized form allows these formulas to be used with a wide range of realistically shaped rotating bodies in general relativity, 
and also with rotating bodies in almost all plausible alternative theories of gravity.
One important advantage of working with the impact parameter 
is that it translates directly into the angular deflection of light, 
which is directly observable by astronomers. Since real astronomical black-hole candidates are believed to be rotating significantly, 
these two rotating impact-parameter formulas can be useful for making comparisons with observations, and may even make it possible to 
determine the rotation parameter.

\end{document}